\documentclass[%
 reprint,
superscriptaddress,
nofootinbib,
amsmath,
amssymb,
aps,
prl,
longbibliography
]{revtex4-2}

\usepackage[version=3]{mhchem} 
\usepackage{lipsum}

\usepackage{longtable}

\usepackage{placeins}
\usepackage{xspace}
\usepackage{appendix}
\usepackage[detect-all]{siunitx}
\usepackage{graphicx}
\usepackage{dcolumn}
\usepackage{bm}
\usepackage{booktabs}
\usepackage{hyperref}
\usepackage{xcolor}
\hypersetup{
    colorlinks,
    linkcolor={blue!70!black},
    citecolor={blue!70!black},
    urlcolor={blue!70!black}
}
\usepackage{dcolumn}
\newcolumntype{L}{D{.}{.}{2,5}}

\newcommand*{\wn}{\ensuremath{\text{cm}^{-1}}\xspace}

\newcommand*{\Pgopher}{\textsc{pgopher}\xspace}

\newcommand*{\eaa}{\ensuremath{\epsilon_{aa}}\xspace}
\newcommand*{\ebb}{\ensuremath{\epsilon_{bb}}\xspace}
\newcommand*{\ecc}{\ensuremath{\epsilon_{cc}}\xspace}
\newcommand*{\ASO}{\ensuremath{\mathcal{A}^\text{SO}}\xspace}

\newcommand*{\XState}{\ensuremath{\widetilde{X}\,{}^2 \! A_1}\xspace}
\newcommand*{\AState}{\ensuremath{\widetilde{A}\,{}^2 \! B_2}\xspace}
\newcommand*{\BState}{\ensuremath{\widetilde{B}\,{}^2 \! B_1}\xspace}

\newcommand\valTB{16353.3787}

\newcommand\valTA{16211.8763}

\newcommand\valAX{0.19153}

\newcommand\valAB{0.19197}

\newcommand\valAA{0.19099}

\newcommand\valBbarX{0.02261}

\newcommand\valBbarB{0.022751}

\newcommand\valBbarA{0.022739}

\newcommand\valBDeltaX{0.00263}

\newcommand\valBDeltaB{0.00270}

\newcommand\valBDeltaA{0.00274}

\newcommand\valDeltaKB{4.5}

\newcommand\valDeltaKA{5.1}

\newcommand\valDeltaJKB{0.19}

\newcommand\valDeltaJKA{1.1}

\newcommand\valDeltaJB{-1.2}

\newcommand\valDeltaJA{-1.9}

\newcommand\valepsaaB{-0.3833}

\newcommand\valepsaaA{0.3791}

\newcommand\valepsbbB{0.0097}

\newcommand\valepsbbA{0.0102}

\newcommand\valepsccB{0.00698}

\newcommand\valepsccA{0.0074}

\newcommand\valDeltasKB{-0.06}

\newcommand\valDeltasKA{2}

\newcommand\nObsBX{538\xspace}
\newcommand\rmsBX{0.0012\xspace}
\newcommand\nObsAX{222\xspace}
\newcommand\rmsAX{0.0023\xspace}

\begin{document}

\title{High-Resolution Laser Spectroscopy of a Functionalized Aromatic Molecule}
\author{Benjamin L. Augenbraun}
\email{augenbraun@g.harvard.edu}
\author{Sean Burchesky}
\author{Andrew Winnicki}
\author{John M. Doyle}
\affiliation{Department of Physics, Harvard University, Cambridge, MA 02138, USA}
\affiliation{Harvard-MIT Center for Ultracold Atoms, Cambridge, MA 02138, USA}

\date{September 13, 2022}

\begin{abstract}
We present a high-resolution laser spectroscopic study of the $\AState - \XState$ and $\BState - \XState$ transitions of calcium (I) phenoxide, CaOPh (\ce{CaOC6H5}). The rotationally resolved band systems are analyzed using an effective Hamiltonian model and are accurately modeled as independent perpendicular ($b$- or $c$-type) transitions. The structure of calcium monophenoxide is compared to previously observed Ca-containing radicals and implications for direct laser cooling are discussed. This work demonstrates that functionalization of aromatic molecules with optical cycling centers can preserve many of the properties needed for laser-based control.
\end{abstract}

\maketitle
The exquisite control over quantum systems achievable using optical photons is a hallmark of modern quantum science and physical chemistry. For example, diatomic molecules controlled at the single-quantum-state level have provided fundamental insights into chemical collisions/reactions~\cite{Cheuk2020,Liu2022} and represent a promising platform for near-term quantum computing architectures~\cite{Kaufman2021, Hudson2018, Ni2018}. Compared to these systems, polyatomic molecules offer qualitatively unique vibrational and rotational motions that enable new opportunities in physics, chemistry, and quantum technology~\cite{Hutzler2020}. All polyatomic molecules have long-lived states with angular momentum arising from nuclear motion, potentially allowing for external-field control of chemical behavior~\cite{Augustoviov2019}, robust encoding of quantum information~\cite{albert2019robust}, and sensitivity to fundamental physical effects~\cite{Kozyryev2021Enhanced, jansen2014perspective, kozyryev2017PolyEDM, Yu2021Probing}.

Science experiments with ultracold atoms and molecules rely on the ability to cool, control, and detect molecules efficiently and, ideally, nondestructively. Optical cycling, a process in which molecules are made to rapidly and repeatedly scatter many hundreds or thousands of photons, is a versatile way to carry out these tasks. However, in general, a molecule that is excited by a narrow-band laser to a single rotational state of an excited electronic state can decay to a range of (metastable) rotational and vibrational states in the electronic ground state, rendering optical cycling a difficult task. Nonetheless, a large class of polyatomic molecules has been identified for which this ``leakage'' is reduced to such an extent that optical cycling and laser cooling might be practical~\cite{kozyryev2016MOR, isaev2015polyatomic, isaev2016laser}. Using these ideas, direct laser cooling of a polyatomic molecule was first achieved in 2016~\cite{kozyryev2016Sisyphus}, followed by rapid progress on more complex molecules such as YbOH~\cite{AugenbraunYbOHSisyphus} and \ce{CaOCH3}~\cite{mitra2020direct}. Recently, the linear radical CaOH has been cooled to the $\mu$K regime, trapped, and loaded into an optical dipole trap where quantum-state control has been realized~\cite{Baum2020, Vilas2022}.

\begin{figure*}[t]
    \includegraphics[width=0.9\textwidth]{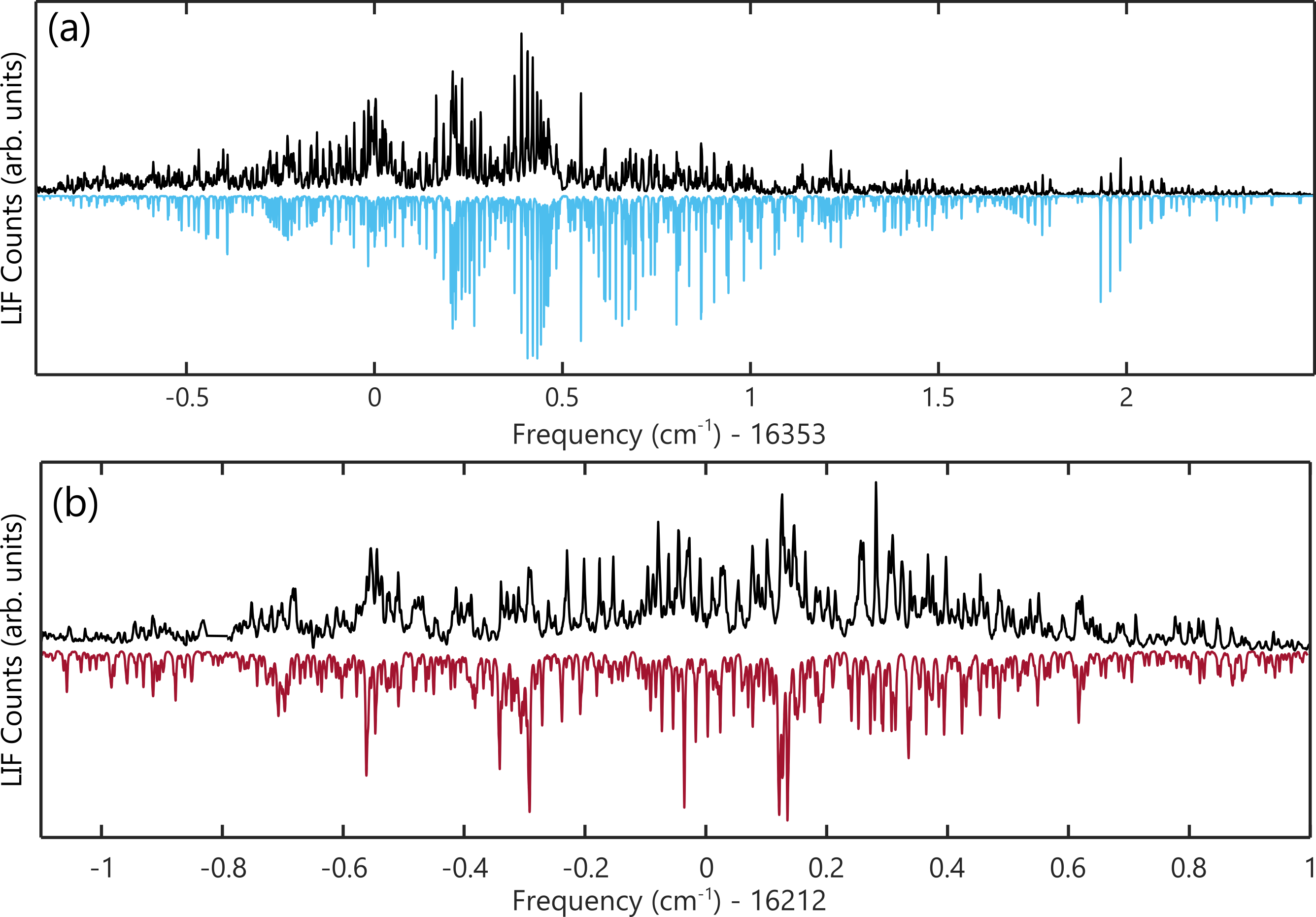}
    \caption{(a) Laser excitation spectrum of the CaOPh $0^0_0 \, \BState \leftarrow \XState$ band near 611.6~nm. The experimental spectrum is plotted as a black, upward-going trace and the optimized simulation is plotted as a blue, downward-going trace. (b) Laser excitation spectrum of the CaOPh $0^0_0 \, \AState \leftarrow \XState$ band near 616.8~nm. The experimental spectrum is plotted as a black, upward-going trace and the optimized simulation is plotted as a red, downward-going trace.}
    \label{fig:LIFFullRange}
\end{figure*}

These experimental demonstrations have motivated intense theoretical efforts to identify new and more complex molecules that can be optically cycled. A concrete method to design molecules that can be controlled by laser light has arisen through analysis of molecules with an MR structure, where M is an alkaline-earth metal atom and R is a pseudohalide (R = F, OH, \ce{OCH3}, SH, etc.). This leads to the idea of an optical cycling center (OCC), which is a quantum functional group that endows a molecule with the capability of optical cycling~\cite{isaev2015polyatomic, kozyryev2016MOR, Ivanov2019Rational, Klos2020, Augenbraun2020ATM, Dickerson2021, Dickerson2021Optical}. Like other functional groups, an OCC ideally modifies the properties of the molecule to which it is attached in a manner that is ``orthogonal'' to the molecule's other properties; the OCC adds the possibility of optical cycling without sacrificing the desirable properties of the molecular fragment that is being modified. While the specific metal-pseudohalide bonding motif appears to be necessary, within this class of molecules OCCs might be realized for a large number of molecules with diverse geometries and constituents. Through the attachment of an OCC to organic ligands of interest, it may be possible to leverage the diversity of molecular structures and symmetries present in Nature for practical applications.

There has been much interest in the specific case of attachment of an OCC to organic molecules containing cyclic hydrocarbons~\cite{Augenbraun2020ATM, Ivanov2020Toward, Dickerson2021, Dickerson2021Optical}. These molecules are interesting due to the importance of cyclic rings in organic chemistry and because the carbon ring represents a highly substitutable platform that can be used to attach, e.g., quantum sensing or quantum information processing nodes. Recently, a set of these molecules---calcium monophenoxide (\ce{CaOC6H5}) and several of its organic derivatives---were produced and detected in the gas phase~\cite{zhu2022functionalizing}. The vibrational branching ratios (VBRs; the probabilities that spontaneous emission is accompanied by a change in the vibrational state) associated with the lowest electronically excited states were found to be favorable for optical cycling. That work strongly motivated the need for rotationally-resolved spectra that provide deeper insight into the molecular structure and that help determine the practical pathway to optical cycling.

Here, we report on the rotationally-resolved spectrum of calcium monophenoxide (CaOPh), a functionalized aromatic molecule in which a phenyl radical is ``decorated'' by a CaO optical cycling center. We study the $0^0_0 \, \AState \leftarrow \XState$ and $0^0_0 \, \BState \leftarrow \XState$ bands using laser excitation spectroscopy to probe molecules in a cryogenic buffer-gas beam. We report rotational and fine-structure parameters for the $v=0$ levels of the \XState, \AState, and \BState states and compare the measured parameters to related Ca-contaning radicals. Finally, we discuss the implications for optical cycling and laser cooling of CaOPh. The measurements presented in this manuscript provide a microscopic view of the rotational and fine-structure parameters for the lowest three electronic states in CaOPh. Comparisons to simpler species (unsubstituted phenols and simpler metal--pseudohalides) further elucidate the orthogonality of the CaO optical cycling center upon substitution to the organic framework. These measurements represent a critical step toward the demonstration of direct laser cooling of molecules containing cyclic hydrocarbons.


To study the rotational and fine structure of CaOPh, we have performed a series of laser-induced fluorescence (LIF) and pump-probe double resonance (PPDR) measurements. Details of the experimental methods are provided in the Supplementary Information. In brief, CaOPh molecules are produced in a cryogenic buffer-gas beam (CBGB) source via ablation of Ca metal in the presence of phenol vapor. The molecules are formed into a collimated molecular beam and probed downstream by monitoring LIF following excitation by a single-frequency dye laser beam. In the PPDR measurements, a second laser beam intersects the molecular beam between the source and detection regions. This laser beam can pump population between rotational states, depleting or enhancing population in a given ground rotational level. The PPDR measurements identify transitions sharing a common level and are useful for quantum number assignment.

CaOPh is predicted to be a planar asymmetric top molecule with $C_{2v}$ symmetry~\cite{Dickerson2021}. 
By analogy with the related species \ce{CaNH2}, the ground $\widetilde{X}$ state is expected to have $A_1$ symmetry while the lowest excited states are \AState (valence electron density oriented in the molecule plane) and \BState (valence electron density oriented perpendicular to the molecule plane)~\cite{Dickerson2021, Whitham1990}. In the linear limit, the \XState term correlates to a ${}^2\Sigma^+$ state and the \AState and \BState terms correlate to the same ${}^2\Pi$ state. Due to the geometrical asymmetry, the \AState and \BState are split and the electronic orbital angular momentum is largely quenched in each of these states~\cite{Marshall2004, Marshall2005, liu2018rotational}.

The rotational energy levels of CaOPh can be labeled by $N_{K_a K_c}$, where $N$ is the rotational angular momentum and $K_a$ and $K_c$ are its projection onto the $a$ and $c$ inertial axes, respectively. The rotational levels are roughly grouped by $K_a$. For given values of $N$ and $K_a$, there are two possible values of $K_c$ ($K_c = N - K_a$ and $K_c = N - K_a + 1$) that are split by the asymmetry doubling. The unpaired electron spin, located primarily on the Ca metal atom, leads to a further splitting of each rotational level due to the spin-rotation interaction, $\mathbf{J} = \mathbf{N} + \mathbf{S}$. While our observations do not have sufficient resolution to observe the effect of nuclear hyperfine structure, proton nuclear spin statistics do affect the relative intensities of observed transitions~\cite{buckingham2013high}. Due to the orientation of the valence electron with respect to the molecular plane, the $\AState-\XState$ transition is expected to obey $b$-type selection rules ($\Delta K_a = \pm 1$ and $\Delta K_c = \pm 1$) while the $\BState-\XState$ transition will obey $c$-type selection rules ($\Delta K_a = \pm 1$ and $\Delta K_c = 0$)~\cite{Gordy1984}.

\begin{figure}[tb]
    \includegraphics[width=1\columnwidth]{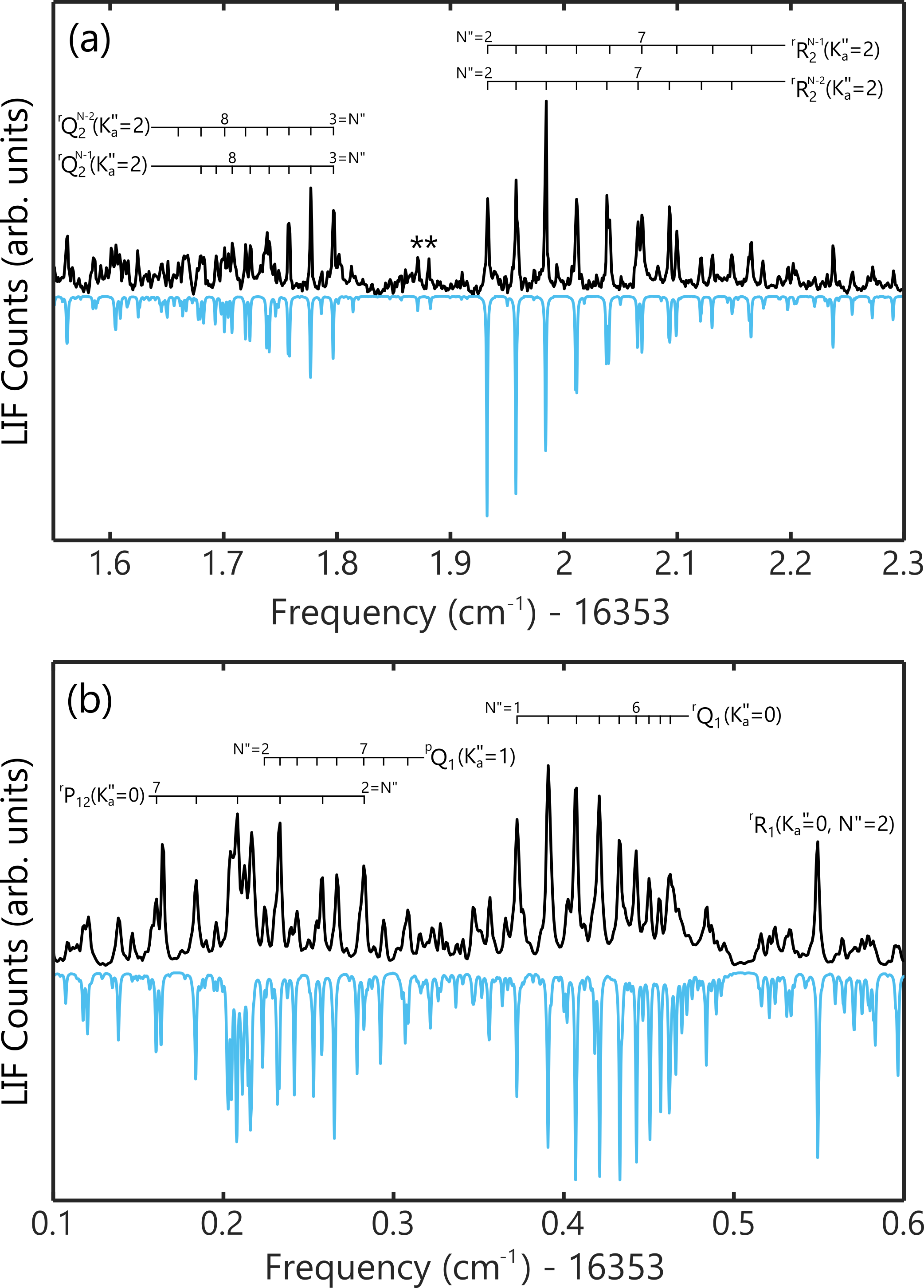}
    \caption{Portions of the observed (black, upward-going) and simulated (blue, downward-going) laser excitation spectrum of the $0^0_0 \, \BState \leftarrow \XState$ band near the origins of the (a) $K_a'=3 - K_a''=2$ and (b) $K_a'=1 - K_a''=0$ / $K_a'=0 - K_a''=1$ subbands. Features marked by $\ast$ in part (a) come from high-$J$ lines of the $^rS_{21}$ subband.}
    \label{fig:BXSubbands}
\end{figure}

The $\BState \leftarrow \XState$ LIF spectrum collected in the range $16352-16355$~\wn is shown in Fig.~\ref{fig:LIFFullRange}(a). The spectrum is very dense due to the small rotational constants ($A \sim 0.2~\wn$, $B\approx C \sim 0.02~\wn$). Analysis began on the $K_a'=3 - K_a''=2$ subband due to its relatively open structure and isolation (see Fig.~\ref{fig:BXSubbands}).\footnote{The branch assignment labels follow the \Pgopher documentation~\cite{western2017pgopher}, and we use the compact notation $^{\Delta K_a} \Delta N_{F_i'',F_j'}^{K_c''}(K_a'')$.} By fitting the R-branch transitions within this subband to approximate analytical values for the asymmetry splittings~\cite{wang1929asymmetrical, Gordy1984}, it was possible to determine that the transition likely had $K_a''=2$ and to estimate values for $(B+C)/2$ and $B-C$. The approximate values of $(B+C)/2$ and $B-C$ were used to construct a Hamiltonian model that loosely matched the observed spectrum.

\begin{figure}[tb]
    \includegraphics[width=1\columnwidth]{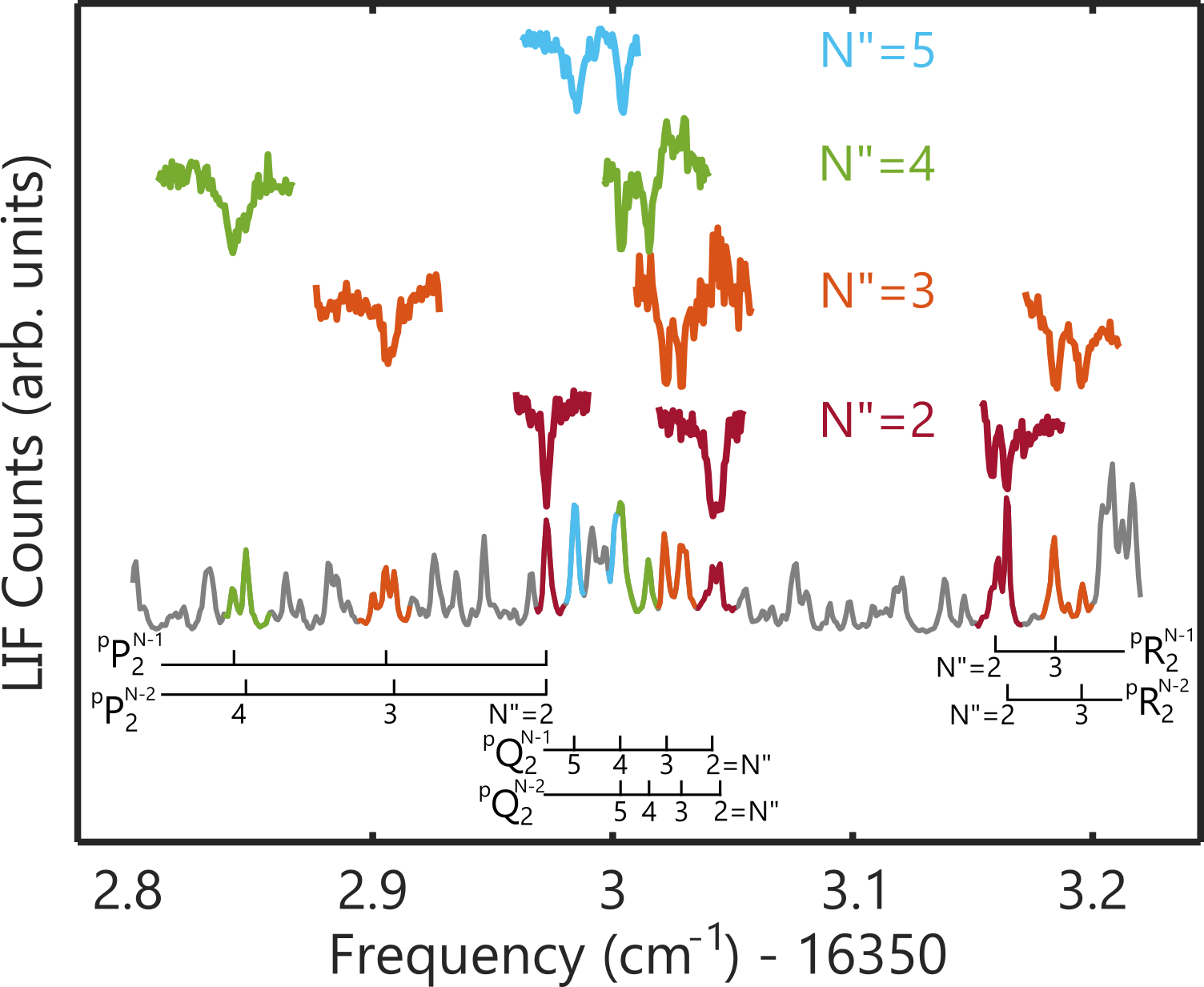}
    \caption{Pump-probe  double resonance (PPDR) spectrum recorded near the origin of the $\BState (K_a'=1) \leftarrow \XState (K_a''=2)$ ($F_2$, upper spin component) subband. Bottom: Raw LIF spectrum recorded while scanning the probe laser over this frequency range (no pump laser present). Top: LIF signals monitored on ${}^rR_2(K_a''=2)$ branch transitions ($N''=2-5$) while the pump laser is scanned over the plotted frequency range. Reduction of probed LIF indicates population was removed from the detected state by the pump laser.}
    \label{fig:DoubleResonance}
\end{figure}

Following rough alignment of the simulated spectrum to the observations, we conducted a series of PPDR measurements to establish quantum number assignments. Typical PPDR are shown in Fig.~\ref{fig:DoubleResonance}. A portion of the ``ordinary" LIF excitation spectrum of the $K_a'=1 - K_a''=2$ subband is reproduced in Fig.~\ref{fig:DoubleResonance}. Above this LIF excitation spectrum are plots showing the dependence of LIF signals originating from $N''=2-5$ of the $K_a'=3-K_a''=2$ subband while the ``pump" laser scans over the distant $K_a'=1 - K_a''=2$ subband. Reduction of the LIF signal is evident whenever the pump laser scans over a transition sharing a common lower level with the transition being monitored. In certain cases, the monitored signal is enhanced due to the pump laser optically pumping population into the rotational state that is detected. In the example of Fig.~\ref{fig:DoubleResonance}, substructure due to excited-state asymmetry splitting is evident. Approximately 30 PPDR scans were recorded and analysis of these spectra made quantum number assignment straightforward.

Assignment of the spectrum was conducted using analysis code adapted from the \Pgopher suite~\cite{western2017pgopher}. The energy levels are modeled using an effective Hamiltonian operator that includes rigid-body rotation, electron spin-rotation coupling, and relevant centrifugal distortion terms,
\begin{equation}
    H_\text{eff} = H_\text{rot} + H_\text{sr} + H_\text{cd,rot} + H_\text{cd,sr}
\end{equation}
In the principal axis system, the rotational Hamiltonian is
\begin{equation}
    H_\text{rot} = A N_a^2 + B N_b^2 + C N_c^2.
\end{equation}
Because CaOPh has $C_{2v}$ symmetry, only the diagonal terms of the electron spin-rotation tensor take nonzero values,
\begin{equation}
    H_\text{sr} = \eaa N_a S_a + \ebb N_b S_b + \ecc N_c S_c.
\end{equation}
The $\Delta^S_K$ centrifugal distortion correction term of the A-reduced form of the spin–rotation interaction was also included, namely~\cite{watson1967determination,brown1979reduced}
\begin{equation}
    H_\text{cd,sr} = \Delta^S_K N_z^3 S_z.
\end{equation}
The form of this operator indicates that it acts like a $K_a'$-dependent spin-orbit interaction.

Molecular constants were optimized using the fitting routines included in \Pgopher. Initial fits were performed by including only those transitions that had been used in the PPDR measurements, due to our confidence in the assignments of these quantum numbers. After achieving satisfactory agreement between the observations and simulated spectra, we included all LIF features with reasonably strong intensity in the fit. The resulting fit (see Tab.~\ref{tab:MolecularConstants} and Supplemental Information) included \nObsBX observations with a root-mean-square (rms) deviation of \rmsBX~\wn. Examination of the line list provided in the Supplemental Information reveals some weak systematic trends in the deviations, which may indicate the need for higher-order distortion constants. Nonetheless, the agreement is quite satisfactory given the congestion of the spectrum and the residuals are commensurate with the measurement uncertainty.

Analysis of the $\AState \leftarrow \XState$ band system (see Fig.~\ref{fig:LIFFullRange}(b)) followed a similar process, although knowledge of the $\XState$ parameters simplified the task. Observations of this band were conducted with higher laser power ($25$~mW), which led to increased power broadening. Initial analysis of this band was conducted by holding the \XState constants fixed to those determined from fits of the $\BState - \XState$ band. The \AState constants were first fixed to those of the \BState, with the exception of $\eaa'$ for which the sign was reversed in accordance with a pure-precession type model~\cite{Morbi1997}. Simulations using these parameters already reproduced the observed spectrum relatively well, allowing us to scan limited portions of the spectrum to conduct PPDR measurements (about 12 features). After confirming quantum number assignments for about a dozen PPDR features, the \AState parameters were allowed to float in our fitting routine. A total of \nObsAX observations were included in this fit, resulting in a rms deviation of \rmsAX~\wn, commensurate with the slightly increased measurement uncertainty for this band.

\begin{table*}[tb]
	\setlength{\tabcolsep}{20pt}
	\begin{center}
		\caption{Spectroscopic parameters$^a$ (in \wn) for CaOPh. The numbers in parentheses represent $2\sigma$ uncertainty estimates. }
		\label{tab:MolecularConstants}
		\begin{tabular}{cccc}
			\hline\hline 
			\multicolumn{1}{c}{Parameter}   & \multicolumn{1}{c}{\XState} & \multicolumn{1}{c}{\AState} & \multicolumn{1}{c}{\BState} \rule{0pt}{2.75ex} \\ \hline
			$T_0$ &   &  $\valTA(4)$ &  $\valTB(2)$ \rule{0pt}{2.5ex} \\
			$A$ & $\valAX(4)$  & $\valAA(14)$ &  $\valAB(6)$  \\
			$(B+C)/2$ &  $\valBbarX(1)$ & $\valBbarA(9)$ & $\valBbarB(10)$  \\
			$B-C$ &  $\valBDeltaX(1)$ & $\valBDeltaA(2)$ &  $\valBDeltaB(2)$  \\
			$\Delta_K \times 10^5$ &  & $\valDeltaKA(7)$ & $\valDeltaKB(2)$  \\
			$\Delta_{JK} \times 10^8 $ &   & $\valDeltaJKA(9)$ & $\valDeltaJKB(4)$   \\
			$\Delta_J \times 10^7$ &  & $\valDeltaJA(6)$ &  $\valDeltaJB(4)$  \\
			$\eaa$ &   & $\valepsaaA(6)$ &  $\valepsaaB(2)$  \\
			$\ebb$ &   & $\valepsbbA(2)$ &  $\valepsbbB(1)$  \\
			$\ecc$ &  & $\valepsccA(2)$ & $\valepsccB(9)$   \\
			$\Delta^{S}_K \times 10^{4}$ &  & $\valDeltasKA(1.5)$ &  $\valDeltasKB(2)$  \\
			\hline\hline
		\end{tabular}
	\end{center}
	\begin{minipage}{1.8\columnwidth}%
		\footnotesize \vspace{-0.65em} $^a$ Parameters represent: $T_0$: band origin, $A,B,C$: rotational constants about principal axes, $\Delta_K, \Delta_{JK}, \Delta_J$: quartic rotational distortion, $\eaa, \ebb, \ecc$: spin-rotation parameters, and $\Delta^S_K$: quartic distortion of the spin-rotation interaction.
	\end{minipage}%
\end{table*}

The molecular constants determined by our measurements provide important insights into the structure of CaOPh and into practical concerns related to optical cycling. The $A$, $B$, and $C$ rotational constants experience only minimal change upon electronic excitation, which is consistent with the observation of highly diagonal Franck-Condon factors reported in the previous low-resolution study of CaOPh~\cite{zhu2022functionalizing}. The measured rotational constants are also reasonable in comparison to related species. The $A$ rotational constants, which characterize rotational about the symmetry axis, are similar to that measured for phenyl radical ($A \approx 0.209~\wn$)~\cite{buckingham2013high} and close to those predicted by \textit{ab initio} theory~\cite{zhu2022functionalizing, Dickerson2021}. The $B$ and $C$ rotational constants are also similar to that expected if one attached CaO to phenyl radical with the Ca--O and O--C bond lengths fixed to the values measured for \ce{CaOCH3}. 

The inertial defect, $\Delta = I_C - I_B - I_A$, is often used as a test of planarity~\cite{oka1995negative}, and planar molecules tend to have small positive values of $\Delta$. Using the rotational constants from Tab.~\ref{tab:MolecularConstants} and assuming $I_B = h/8\pi c^2 B$ (and similarly for $I_A$ and $I_C$), we estimate $\Delta_{\tilde{X}} = -0.995(3) \text{ amu \AA}^2$. 
It is interesting that the inertial defect is negative despite the expectation that CaOPh is planar. Several factors may explain this. First, the rotational constants are affected by second-order corrections due to spin-orbit interactions that likely reduce the $A$ rotational constant, thereby lowering $\Delta$~\cite{Steimle2002Rotational}. Second, low-frequency out-of-plane bending modes are associated with negative inertial defects~\cite{oka1995negative}, and CaOPh has an out-of-plane bending mode at $\nu \approx 45$~\wn~\cite{zhu2022functionalizing}. The empirical relation of Oka~\cite{oka1995negative} predicts a value $\Delta \approx -0.55 \text{ amu \AA}^2$ when considering just this one out-of-plane bending mode. This suggests that the negative inertial defect does not contradict a planar equilibrium geometry, although future isotopic studies of CaOPh are necessary to fully determine the molecular geometry.

The dominant excited-state fine structure parameter is the spin-rotation constant $\eaa$. The effective Hamiltonian term associated with this parameter acts like a rotation-dependent spin-orbit splitting due to its form, $H \propto \eaa K_a \Sigma$. The interpretation of \eaa in terms of second-order spin-orbit effects is well developed~\cite{Whitham1990,Marr1995a, Morbi1997}. Using perturbation theory and the pure-precession approximation, the second-order contribution due to mixing between the $\tilde{A}$ and $\tilde{B}$ states is given by~\cite{Morbi1997}
\begin{equation}
    \eaa^{(2)} \approx \pm \frac{4 A \ASO}{E_{\tilde{B}} - E_{\tilde{A}}},
    \label{eq:eaaSecondOrder}
\end{equation}
where the spin-orbit constant \ASO can be approximated from the spin-orbit splitting of a related linear molecule (e.g., CaOH). In Eq.~\ref{eq:eaaSecondOrder}, the positive (negative) sign applies to the \AState (\BState) state. Taking $\ASO \approx 67~\wn$~\cite{Li1992} leads to the prediction $\eaa^{\tilde{A}/\tilde{B}} \approx \pm 0.36~\wn$, in excellent agreement with the observed values $\eaa^{\tilde{A}} = \valepsaaA~\wn$ and $\eaa^{\tilde{B}} = \valepsaaB~\wn$. The second-order contribution accounts for essentially the entire value of the measured \eaa constants. The agreement between theory and experiment suggests that, to a good approximation, the \AState and \BState states interact only with each other and both correlate to a ${}^2\Pi$ state in the linear limit. The validity of simple perturbation theory expansions suggest that the valence electron residing on the CaO moiety is largely unaffected by the complex organic ligand, validating the OCC concept.

A simple two-level model can be used to shed further light on coupling between the \AState and \BState states due to the spin-orbit interaction~\cite{liu2018rotational}. Consider a basis comprising the components of a $p\pi$ orbital, $\lbrace \lvert A' \rangle, \lvert A'' \rangle \rbrace$, where $A'$/$A''$ refers to the reflection symmetry with respect to the molecule plane. In this basis, we can write~\cite{liu2018rotational}
\begin{equation}
    H = \frac{1}{2} \begin{pmatrix}
        \Delta E_0 & i \ASO \\
        -i \ASO & -\Delta E_0
    \end{pmatrix},
    \label{eq:HamiltonianTwoLevel}
\end{equation}
where $\Delta E_0$ is the $\tilde{A}-\tilde{B}$ splitting in the absence of spin-orbit interactions. Equation~\ref{eq:HamiltonianTwoLevel} leads to an energy separation between the two coupled states that is
\begin{equation}
    \Delta E = \sqrt{(\ASO)^2 + (\Delta E_0)^2} \, .
\end{equation}
We have measured $\Delta E \approx 142~\wn$ and thus $\Delta E_0 \approx 123~\wn$. From Eq.~\ref{eq:HamiltonianTwoLevel}, each of the \AState and \BState states is approximately 95\% pure. This purity indicates that the electronic orbital angular momentum in each state is largely quenched due to the presence of off-axis atoms~\cite{Marshall2004, Marshall2005}. For comparison, in \ce{CaNH2} the analogous states are more than 99\% pure~\cite{Morbi1997, Morbi1998}, i.e. in CaOPh the extent of orbital angular momentum quenching is slightly less than in \ce{CaNH2}.

An important result of this work for the purposes of laser cooling is the identification of rotational transitions associated with closed optical cycling transitions. For the CaOPh $0^0_0 \, \BState \leftarrow \XState$ band, optical cycling primarily makes use of $c$-type ($\Delta K_a = -1$) transitions originating from $N_{K_a K_c} = 1_{10}$ rotational level of $\XState$~\cite{Augenbraun2020ATM}. These transitions have been identified (see Supplemental Information) and can now be used in attempts to observe optical cycling. Mixing between the \AState and \BState states at the 5\% level indicates that additional sidebands may be necessary on the laser beams used for optical cycling, similar to the case of \ce{CaOCH3}~\cite{mitra2020direct}. Interestingly, in the case of CaOPh this arises from the orbital angular momentum induced by mixing of electronic states, while in \ce{CaOCH3} the need for a rotational repumper is expected based on standard electric dipole selection rules. Rotational repumping is not likely to be a major experimental impediment because the small rotation and spin-rotation constants mean that these additional frequencies can be added using readily available acousto- or electro-optic modulators. 

In summary, the $0^0_0 \, \AState \leftarrow \XState$ and $0^0_0 \, \BState \leftarrow \XState$ transitions of CaOPh have been observed at high resolution and rotationally analyzed. For each band, we have found good agreement between observations and a theoretical model without the need to include coupling among the closely-spaced excited states. Our measurements yield rotational and fine structure constants that agree well with theoretical expectations. Due to spin-orbit and vibronic interactions that couple the $\tilde{A}$ and $\tilde{B}$ states, this pair of states represents an interesting platform for future detailed studies of the (pseudo)-Jahn-Teller effect~\cite{liu2018rotational, melnik2007development}. Our measurements provide the rotational frequencies needed to attempt optical cycling of CaOPh. Based on previous measurements of the VBRs~\cite{zhu2022functionalizing}, vibrational leakages will limit the number of photon scatters to $\sim$20 photons per molecule without vibrational repumping lasers. Future studies of the vibrationally excited levels in \XState are needed to locate these repumping pathways and extend the optical cycling to $>$100 photons per molecule, a level that is sufficient for transverse cooling~\cite{kozyryev2016Sisyphus}. The results presented here provide the laser information required to carry out these efforts and to study whether optical cycling can be realized in large, complex molecules like CaOPh.

\textit{Supplemental Information ---} Details about the experimental setup, data acquisition, and analysis procedure can be found in the supplemental information. Also provided are line lists for the $\AState - \XState$ and $\BState - \XState$ origin bands of CaOPh.


\section{Acknowledgments}
The authors thank Profs. Eric Hudson, Wes Campbell, and their groups, for stimulating discussions during this work. The authors are also grateful to Prof. Timothy Steimle for his valuable insights about the CaOPh spectra and Prof. Jinjun Liu for useful discussions related to coupling between the \AState and \BState states.  This work was supported by the W. M. Keck Foundation, the Heising Simons Foundations, and the AFOSR. SB acknowledges financial support from the NSF GRFP. 

\bibliography{CaOPh_library}

\end{document}